\documentclass{llncs}
\usepackage{hyperref}
\usepackage{subcaption}
\usepackage{graphicx}

\begin{document}
\title{Prototype tests for a highly granular scintillator-based hadronic calorimeter}

\author{Yong Liu, for the CALICE Collaboration}

\institute{Institut f\"ur Physik and Cluster of Excellence PRISMA, 
\\Johannes Gutenberg-Universit\"at Mainz, 55099 Mainz, Germany \\
\email Email: yong.liu@uni-mainz.de}

\maketitle

\begin{abstract}
Within the CALICE collaboration, several concepts for the hadronic calorimeter of a future lepton collider detector are studied. After having demonstrated the capabilities of the measurement methods in "physics prototypes", the focus now lies on improving their implementation in "technological prototypes", that are scalable to the full linear collider detector. The Analogue Hadronic Calorimeter (AHCAL) concept is a sampling calorimeter of tungsten or steel absorber plates and plastic scintillator tiles read out by silicon photomultipliers (SiPMs) as active components. The front-end electronics is fully integrated into the active layers of the calorimeter and is designed for minimal power consumption (i.e. power pulsing). The versatile electronics enables the prototype to be equipped with different types of scintillator tiles and SiPMs. In recent beam tests, a prototype with $\sim$3700 channels, equipped with several types of scintillator tiles and SiPMs, was exposed to electron, muon and hadron beams. The experience of these beam tests resulted in an optimal detector design with surface-mounted SiPMs suitable for the automated mass assembly. The proceeding will cover topics including the testbeam measurements with the AHCAL technological prototype, the improved detector design and the ongoing development of a large prototype for hadronic showers.
\end{abstract}

\section{Introduction}
Precision physics programs at future lepton colliders impose stringent requirements on the calorimeter performance. In order to reach an unprecedented jet energy resolution. calorimeters are required to be highly granular and compact inside the magnet coil. The CALICE collaboration has been developing highly granular options for electromagnetic and hadronic calorimeters~\cite{Overview}.

The Analogue Hadronic Calorimeter (AHCAL), as one of the CALICE technical options, is a sampling hadron calorimeter concept based on scintillator tiles ($\mathrm{30\times30\times3~mm^{3}}$) coupled to silicon photomultipliers (SiPMs) as active components. Iron and tungsten have been used as absorber materials. A large physics prototype has been exposed to various beams and the capabilities of the AHCAL concept have been demonstrated~\cite{PhysProt}. The current focus lies on the implementation of a large technological prototype, which needs to be scalable to a full detector at a future lepton collider.

\section{AHCAL beam-test campaigns and highlights}

\begin{figure}
\centering 
\includegraphics[width=.8\textwidth]{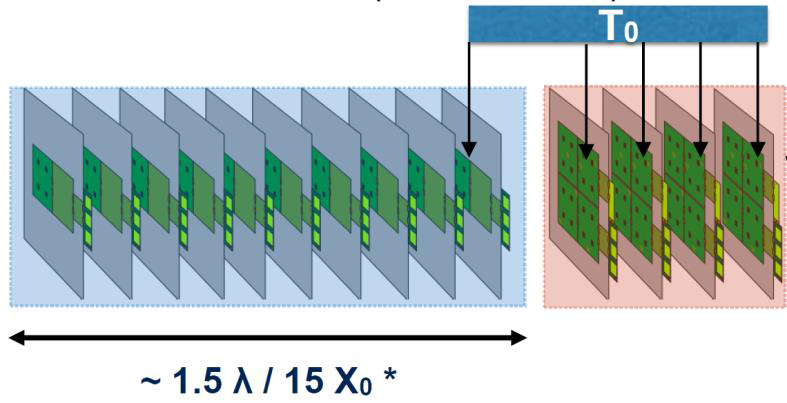}
\caption{The schematics of the AHCAL prototype at CERN SPS in 2015}
\label{fig:Setup}
\end{figure}

An AHCAL prototype with 14 active layers (3744 channels in total) was tested using beams at CERN SPS in 2015 using a steel stack and a tungsten stack, respectively, as shown in Fig.~\ref{fig:Setup}. Large data sets have been collected, including muons for the detector calibration, electrons for precision electromagnetic shower studies and pions for detailed hadronic shower studies.

The SiPM response to single photons has been obtained using the UV light generated by an on-board LED for each channel. The gain of each SiPM was extracted from the LED data, which can also monitor temperature variations. The detector was calibrated using muon beams as minimum ionising particles (MIPs). 

The readout electronics in the AHCAL prototype with a feature of fine timing measurements enables possibilities to study the temporal evolution of hadronic showers. The reference time (T0) signals come from trigger scintillator plates placed in front and rear of the AHCAL detector and the time of hits (T) in each active layer can be compared to T0. Preliminary results of muon data show similar distributions of the time difference T-T0 in steel and tungsten absorbers. Based on the time calibration procedure established by muon data, the timing analysis of electron and pion data is ongoing.

\section{AHCAL mass assembly}
\begin{figure}[htbp]
\centering 
\includegraphics[width=.45\textwidth]{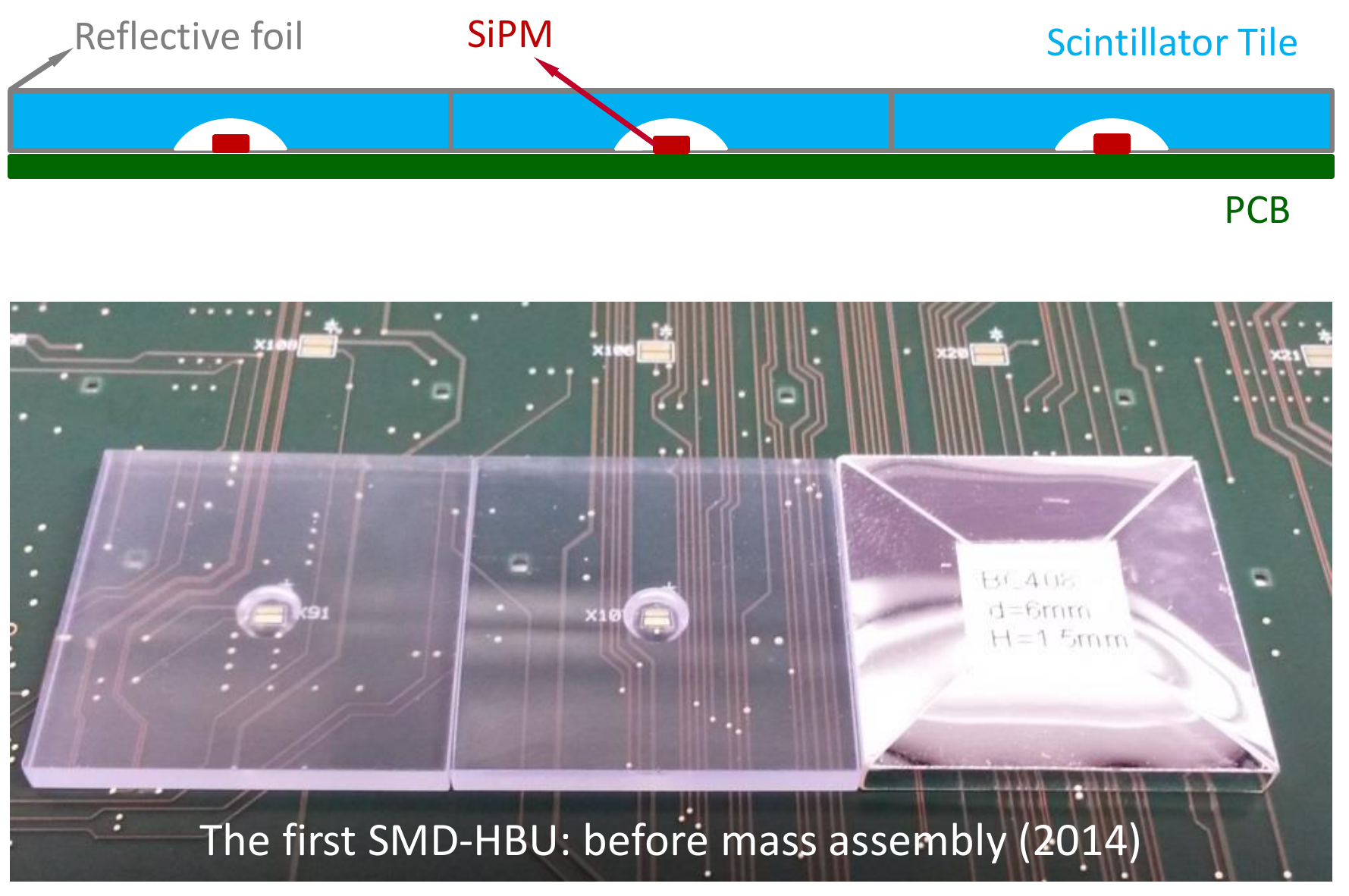}
\includegraphics[width=.45\textwidth]{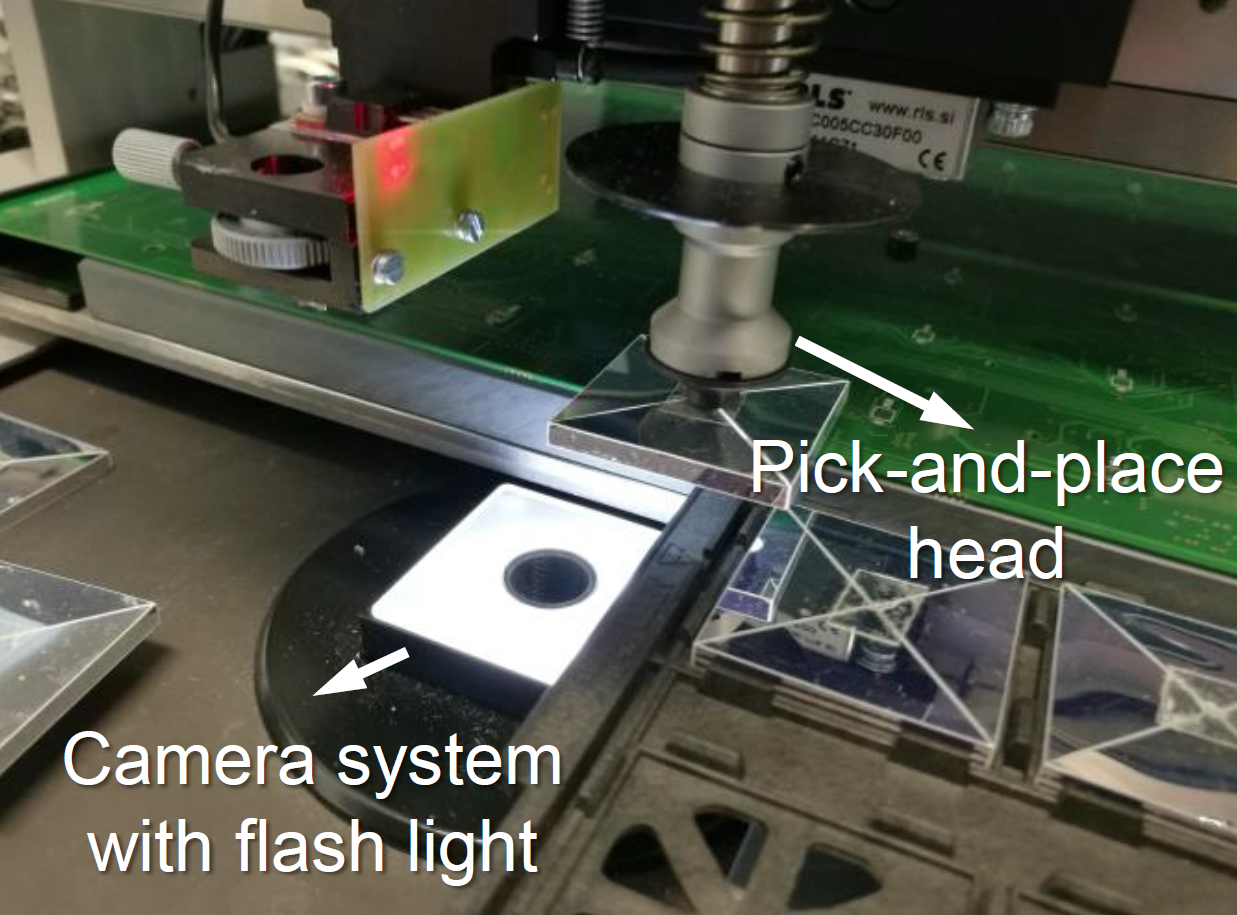}
\caption{Schematics of the surface-mounted design (top left), the first SMD-HBU (bottom left) and a moment of the automated tile assembly (right)}
\label{fig:SMD-HBU}
\end{figure}

Various designs for the SiPM-tile coupling have been tried in the AHCAL prototype in the CERN beamtests. The surface-mounted design turned out to be the only solution which is suitable for mass assembly. In this design, a surface-mounted SiPM (SMD-SiPM), soldered on the PCB, directly couples to from top to a tile with a dome-shaped cavity. The thin package is fully accommodated inside the cavity, as shown in Fig.~\ref{fig:SMD-HBU}. The design was optimised by the GEANT4 full simulation to achieve a high and uniform light collection efficiency when particles pass through different positions~\cite{Uniformity}.

\begin{figure}[htbp]
\centering 
\includegraphics[width=.45\textwidth]{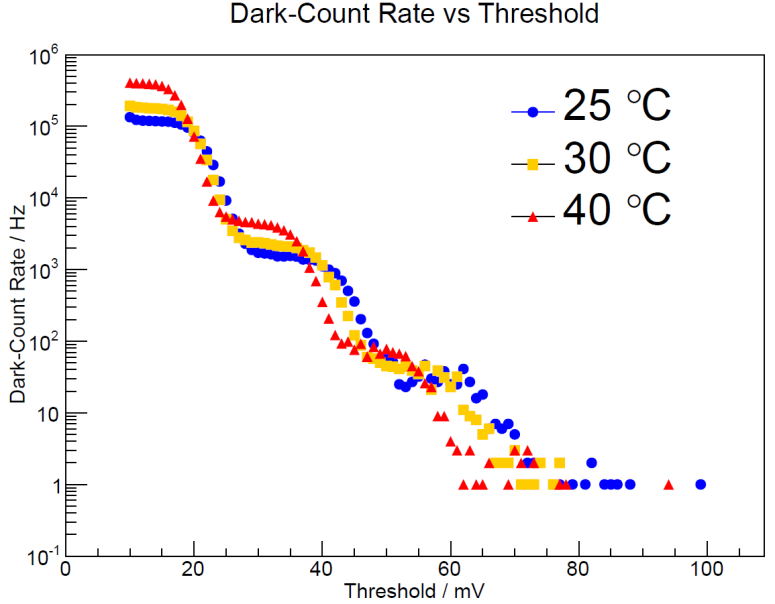}
\includegraphics[width=.45\textwidth]{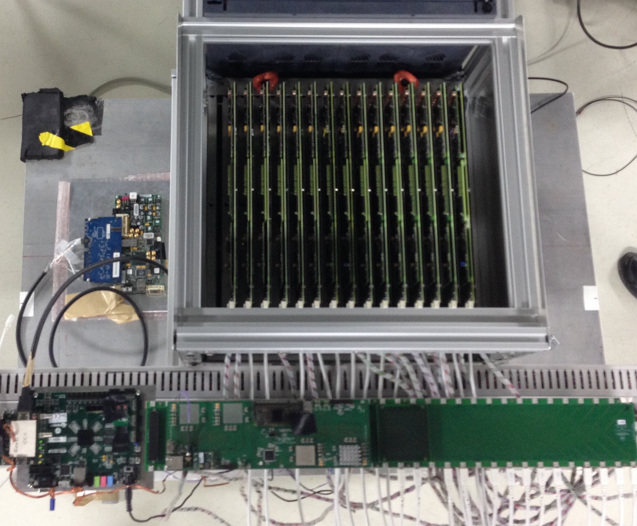}
\caption{Measurements of dark-count rates at different thresholds with a new low-crosstalk SiPM (left); a new steel stack with 15 active layers (right)}
\label{fig:SiPM_NewStack}
\end{figure}

A first proof-of-principle readout board with 144 channels using this design (SMD-HBU) was successfully built in 2014 via mass assembly using a pick-and-place machine with electronics established for SMD-SiPMs and individually wrapped tiles. In 2016, six new SMD-HBUs were built using new SiPMs and an updated tile design while the mass assembly routine was fully practised (Fig.~\ref{fig:SMD-HBU}). With the benefit of new SiPMs, dark-count noise is reduced and inter-pixel crosstalk is dramatically suppressed. One such new SiPM was characterised at different temperatures. Fig.~\ref{fig:SiPM_NewStack} shows the crosstalk level is lower than 2~\% at the nominal reversed voltage. Also the uniformity of the SiPM quality, piece by piece as well as pixel by pixel within the same device, is much improved.

\section{AHCAL technological prototype}
As the first step towards developing a large AHCAL technological prototype, a small prototype of 15 active layers with a single HBU per layer has been developed using high-quality SiPMs for electromagnetic shower studies.

Interface boards of 15 layers have been updated. The detector interface (DIF) board have been redesigned, equipped with a modern FPGA. The power board has been improved for better power-pulsing performance and for lower heat dissipation. Meanwhile, new HBUs have been developed for the updated ASIC chips (SPIROC2E) in a new package (BGA). Preliminary LED tests show that there is no observable gain drop with a switch-on time of 60 $\mathrm{\mu s}$ in the power-pulsing mode, compared to the constantly running mode.

The small prototype was tested using electron beams at DESY for precision measurements of electromagnetic showers and for testing the power-pulsing performance. Promising performance has been achieved: all channels are working and the channel-wise SiPM gain is quite uniform. Efforts are being invested to check the performance differences between with and without power pulsing and to apply corrections to the SiPM saturation effect and temperature variations.

The ultimate goal of the AHCAL technological prototype is to instrument 40 active layers with $2\times2$ HBUs per layer in a steel stack ($\sim1~\%$ of the barrel ILC-AHCAL) to demonstrate the scalability to build a final detector. 

Steady steps have been made towards the mass production and quality assurance. More than 24000 tiles have been produced via injection moulding. Dedicated test stands have been developed to check the functionalities of readout chips, to fully characterise SiPM samples and to test fully assembled HBUs.

\section{Summary and outlook}
The scintillator-based hadronic calorimeter concept is being developed within the CALICE collaboration. Based on the experience from successful beam tests for several AHCAL prototypes, an optimal detector design suitable for the mass assembly has been chosen and promising performance has been achieved. In addition, procedures for the automated assembly and testing have been established.

The AHCAL technological prototype is scheduled to be built within 2017. Meanwhile, there will be a beam test with a strong magnetic field in 2017. More tests by exposing the large prototype to hadron beams are expected in 2018.

\end{document}